\documentclass[aps,twocolumn,pre]{revtex4}

\pdfoutput=1

\usepackage[utf8]{inputenc}
\usepackage{lmodern}
\usepackage{hyperref}
\usepackage{amsmath}
\usepackage{amssymb}
\usepackage{graphicx}
\usepackage{xcolor}
\usepackage{color}
\usepackage{amssymb}
\usepackage{amsthm}
\usepackage{epstopdf} 
\usepackage{natbib}
\usepackage{dcolumn}
\usepackage{amsfonts}
\usepackage{bm}

\newcommand{\mic}{\,\mu\mathrm{m}}
\newcommand{\nm}{\,\mathrm{nm}}

\newcommand{\kev}{\,\mathrm{keV}}
\newcommand{\fs}{\,\mathrm{fs}}

\begin{document}

\title{On the physics of electron ejection from laser-irradiated overdense plasmas}

\author{M. Th\'evenet, H. Vincenti, J. Faure}
\address{LOA, ENSTA ParisTech, CNRS, Ecole polytechnique, Universit\'e Paris-Saclay, 828 bd des Mar\'echaux, 91762 Palaiseau cedex France}

\begin{abstract}

Using 1D and 2D PIC simulations, we describe and model the backward ejection of electron bunches when a laser pulse reflects off an overdense plasma with a sharp density gradient on its front side. The dependence on the laser intensity and gradient scale length is studied. It is found that during each laser period the incident laser pulse generates a large charge-separation field, or plasma capacitor, which accelerates an attosecond bunch of electrons towards vacuum. This process is maximized for short gradient scale lengths and collapses when the gradient scale length is comparable to the laser wavelength. We develop a model that reproduces the electron dynamics and the dependence on laser intensity and gradient scale length. This process is shown to be strongly linked with high-harmonics generation via the Relativistic Oscillating Mirror mechanism.

\end{abstract}

\maketitle

\section{introduction}

The interaction of ultra-high intensity lasers with overdense plasmas is of particular importance for applications such as fast ignition for inertial confinement fusion~\cite{Tabak,Baton2003}, ion acceleration~\cite{Maksimchuk2000,Chen2001a,Yuan2008a} and high-harmonic generation. In all these fields, electrons accelerated at the front surface are the main pathway for energy transfer between the laser and the plasma, and they also drive most of the physical phenomena of interest. High-Harmonic Generation (HHG) in this regime is described by the Relativistic Oscillating Mirror (ROM): electrons from the front surface of the solid target form a very dense sheet of electrons which is driven by the intense laser pulse and radiates high harmonics. This interaction offers the opportunity to generate intense attosecond bursts of X-UV radiation which can then be used as a unique tool to probe matter on extreme scales~\cite{Drescher2002,Uiberacker2007,Horvath2007,Goulielmakis2010}.

Harmonic emission via the ROM mechanism is efficient in the \textit{plasma mirror} regime, \text{i.e.} when the laser impinges the target at oblique incidence in p-polarization on an overdense plasma with a very sharp density gradient. Experiments have shown that the emission is efficient when the gradient scale length is around $L\simeq \lambda_0/10$, where $\lambda_0$ is the laser wavelength~\cite{Kahaly2013}. Such solid density plasmas with very sharp gradients are referred to as \textit{plasma mirrors} because they essentially behave like mirrors that reflect the incident laser pulse with high reflectivity~\cite{Thaury2010} and little spatial deformation.

In parallel to HHG, many experiments have measured electrons emitted in the backward direction, \textit{i.e. ejected} from the front surface of the target toward vacuum. These experiments usually produce large divergence electron beams with energies ranging from hundreds of keV~\cite{Kodama2000,Chen2001a,Li2003,Cai2004a} to few MeV~\cite{Mordovanakis2009}. Some experiments reported on beams emitted in the specular direction~\cite{Mordovanakis2009}, while others reported on beams emitted along the surface~\cite{Chen2006} or in the normal direction~\cite{Li2001}. The diversity in the results can be explained by the fact that these experiments used different gradient scale lengths and intensities. Experiments with a controlled gradient~\cite{Mordovanakis2009,Thevenet2015} showed that bright and well defined electron beams were obtained preferentially for small gradients, typically $L\simeq\lambda_0/10$. Evidently, the gradient scale length is a crucial parameter as it determines the dominant laser absorption mechanism: $J \times B$ heating \cite{Kruer1985} vacuum heating~\cite{Brunel1987} at short $L$, resonant absorption ~\cite{Forslun1975} at intermediate $L$, or parametric instabilities at longer $L$.

Although efficient HHG via the ROM mechanism and electron ejection seem to occur in the same parameter regime, there has never been an effort to understand these emissions as part of the same mechanism. In addition, electron ejection from the front surface is not well understood: several numerical studies report on the subject~\cite{Ruhl1999,Sheng2000,Naumova2004a,Geindre2010} but so far, no theory has been developed in order to explain and predict electron ejection. Some models have been developed in the context of HHG in order to describe the dynamics of the plasma surface~\cite{Baeva2011,Gonoskov2011}, but these models are not useful for describing electron emission because of the strong hypothesis that they rely upon.

The goal of this article is to investigate the mechanism of electron ejection for a large range of parameters and to relate it to high harmonic generation. We focus on the case of plasma mirrors, \textit{i.e.} sharp density gradients. In section \ref{sec1D}, we present a comprehensive numerical study using 1D Particle In Cell (PIC) simulations. These results are used to elaborate a scenario of electron ejection and acceleration, which is described in section \ref{secscenario}. In section \ref{secmodel}, we present a 1D model which highlights the main ingredients that are responsible for electron ejection. The model agrees well with the results of the 1D PIC simulations. Finally, section \ref{sec2D} extends the study to 2D PIC simulations which confirm the 1D study. It is found that HHG and electron ejection are correlated at short gradient scale lengths.

\begin{figure}
\includegraphics[width=\columnwidth]{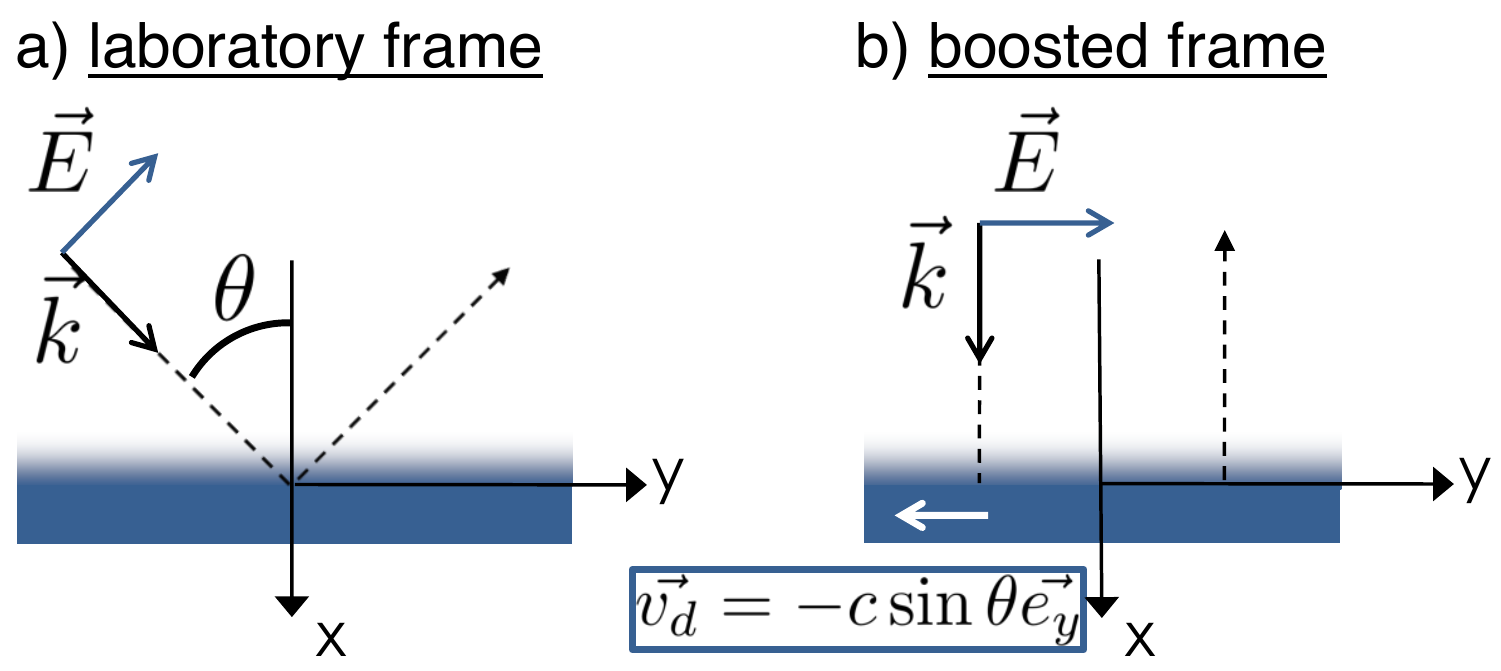}
\caption{Geometry of the interaction -- a) in the lab frame -- b) in the boosted frame obtained after a Lorentz transform at velocity $\mathbf{v}=c\sin\theta\mathbf{e_y}$.}
\label{schema}
\end{figure}

\begin{figure*}[th!]
\includegraphics[width=\textwidth]{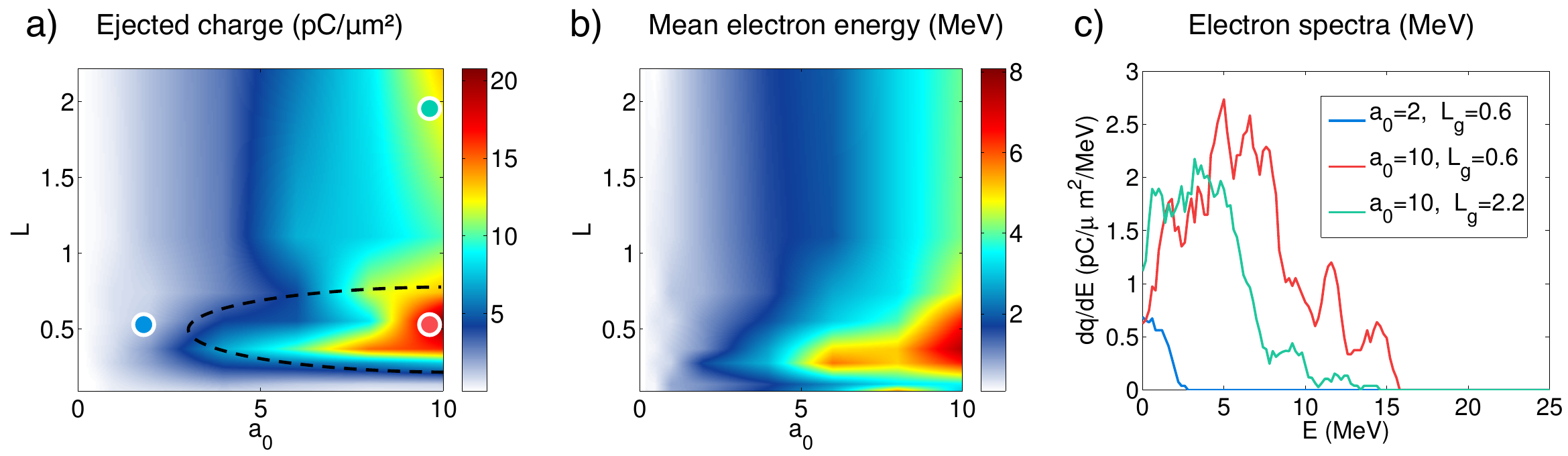}
\caption{Results of a 1D PIC simulation scan ($a_0$,$L$) -- a) Ejected charge. Particles were detected at a distance $d=30$ from the target, and the charge was integrated all along the simulation. -- b) Mean energy of the ejected electrons -- c) Electron spectra from 3 simulations with parameters indicated by the color circles in panel a).}
\label{scan}
\end{figure*}

\section{Parametric study using 1D PIC Simulations}\label{sec1D}

The geometry of the interaction and the notations are depicted in figure \ref{schema}a): $x$ is the direction normal to the target and $y$ is the direction parallel to the target. The electric field of the laser pulse $\mathbf{E_L}$ and its wave vector $\mathbf{k}$ are in the $x-y$ plane, while the laser magnetic field is parallel to the $z$-axis. The laser field impinges the surface with an incident angle $\theta$. This problem is inherently 2D, even in the case where the laser is modeled by a plane wave. 

In order to transform this 2D problem into a 1D problem, we move to the boosted frame introduced by Bourdier~\cite{Bourdier1983}, as illustrated in figure \ref{schema}b). This new reference frame $S'$ is obtained by performing a Lorentz transform to a frame moving at $\mathbf{v}=c\sin\theta \mathbf{e_y}$, with $c$ the velocity of light. Therefore, in the boosted frame, the plasma is no longer at rest but drifts with velocity $\bm{v_d}=-c\sin\theta\bm{e_y}$. In addition, it is straightforward to show that the wave vector is now perpendicular to the surface $\bm{k_0'} = k_0'\bm{e_x}$ and the laser electric field is along the $y$-axis: $\mathbf{E_L'}=E_0\cos\theta\mathbf{e_y}$, where $E_0$ is the laser amplitude in the laboratory reference frame. In the boosted frame, this problem is now purely 1D and can be studied extensively using 1D PIC simulations. In addition, it permits to decouple the laser electric field $E_L$, which is along $y$, from the plasma electric fields, $E_p$ which can only be along $x$.

In the boosted frame, the physical quantities are modified according to the Lorentz transformation. For the laser frequency, lengths along $y$ and plasma density, this reads
\[
\begin{array}{ccc}
 \omega_0' = \omega_0\cos\theta,  & dy'=dy\cos\theta, &n'=n/\cos\theta 
\end{array}
\]
Accordingly, the laser electric field and laser magnetic field amplitude transform to
\[
\begin{array}{cc}
E'_0=E_0\cos\theta & B'_0=E_0\cos\theta/c
\end{array}
\]
 We now adopt the following normalization scheme: 
 \[
\begin{array}{rclcrcl}
t' & \equiv & \omega_0' t', & & x' & \equiv & k_0'x' \\
v' & \equiv & v'/c, & & p' & \equiv & p'/m_ec
\end{array}
\]
 where $m_e$ is the electron mass. All fields are normalized as follows
 \[
\begin{array}{rclcrcl}
E' & \equiv & \frac{eE'}{m_e \omega_0' c}, & B' & \equiv & \frac{eB'}{m_e \omega_0'} \\
\end{array}
\]
where $e$ is the electron charge. Therefore, in the boosted frame, the amplitude of the laser field is simply the normalized vector potential $a_0'=eE_0'/m_e \omega_0' c=a_0$. Finally, we choose to normalize the density by the critical density in the laboratory frame $n_c=\epsilon_0m_e\omega_0^2/e^2$ (as opposed to $n'_c$ in the boosted frame): $n' \equiv n'/n_c$. In the following, all quantities are considered in the boosted frame, except if specified otherwise, and for the sake of clarity the prime symbols will be skipped. 
 
The main parameters affecting the emission of backward electrons are (i) the laser amplitude $a_0$, (ii) the plasma density gradient scale length $L$, (iii) the angle of incidence $\theta$ and (iv) the pulse duration. In this study, we will restrict ourselves to the case of femtosecond pulse durations, typically 20-30 fs, as used in most current experiments. In this case, ion motion does not play a significant role during the interaction. In order to study the role of the main parameters $a_0$ and $L$, we performed a set of one hundred 1D Particle-In-Cell (PIC) simulations, which require little computer resources. 

In the laboratory frame, a $800\nm$, $25\fs$ laser pulse impinges on a solid-density plasma with angle of incidence $\theta=45^{\circ}$. Its amplitude is varied from $a_0=0.2$ to $a_0=10$. The plasma bulk density is $250n_c$ and the gradient length is varied from $\lambda_0/100$ to $\lambda_0$ (i.e. in normalized units $L=0.04$ to $L=4.4$). We assume the density gradient to have an exponential shape $n(x)=n_ce^{x/L}$, so that at $x=0$, the electron plasma density reaches $n_c$. The density gradient is artificially cut at the plasma edge $x_e$ defined as $n(x_e)=1/5$. The simulations were performed in the boosted frame, hence numerical conditions are given in this frame: the numerical space-step was $\delta x=1/700$, and we used $1000$ particles-per-cell. The simulation box was $\Delta x=130$ large. Ions were mobile (we took oxygen ions to represent the lightest ions in a Silica target) but simulations with immobile ions yielded very similar results

When simulating the ejection of electrons with 1D simulations, two effects must be considered. First, the laser does not diffract away, so that the laser intensity is greatly overestimated as soon as the propagation distance is larger than a Rayleigh length. Second, charged particles are represented by charged surfaces. Therefore, the electrostatic force between two charged particles does not depend on the distance $r$ between them, while it decreases in $1/r^2$ in a 3D geometry. As electrons leave the target, the plasma surface becomes positively charged and exerts a recall force that does not depend on the electron position. Therefore, if one runs a 1D simulation long enough, all electrons will eventually return to the plasma and the ejected charge will always tend toward zero. In order to obtain realistic results, we chose to consider that electrons are ejected if they cross a plane located at $d=30$ from the plasma edge (i.e. at $7\lambda_0$ from the plasma edge). This distance was chosen (i) much smaller than the Rayleigh length of most current experiments, so that the 1D approximation remains valid, and (ii) much larger than the gradient lengths we studied, so that electrons are considered to be detected far from the plasma surface. This point is quite crucial as detecting electrons too close to the plasma surface will considerably overestimate the ejected charge, while detecting them too far leads to wrong results due to the invalidity of the 1D approximation.

Figure~\ref{scan}a) shows the ejected charge as a function of $a_0$ and $L$. First, it is clear that there are no ejected electrons when the gradient scale length is $L=0$. This has been explained by the gyromagnetic effect~\cite{Geindre2006}: with $L=0$, the plasma behaves as a perfect conductor. The magnetic component of the incident an reflected wave are added together, resulting in a strong magnetic field that prevents electrons from escaping the plasma. This effect is neutralized with longer gradient. For a given value of $a_0$, the ejected charge increases with $L$, reaches a maximum for $L=L_{max}\simeq0.4$ (i.e. $\sim \lambda_0/10$), and then slowly decreases. We find that the value of the optimum gradient $L_{max}$ depends weakly on $a_0$ in this range. Evidently, the results also show that the ejected charge increases with the laser amplitude $a_0$. In fig.~\ref{scan}b), the mean energy of ejected electrons is plotted in the same parameter space. It varies in a similar fashion as the ejected charge: the higher the ejected charge, the more energetic the electrons. The electron spectrum is plotted on panel c), for three different simulations represented by the color circles in fig.~\ref{scan}a). The red and green curve stand for $a_0=8$ and a gradient length respectively $L\simeq L_{max}$ and $L \gg L_{max}$. The electron spectra are quite broad and electron energies are in the few-MeV to 10 MeV range. 

These findings agree qualitatively with recent experiment results \cite{Mordovanakis2009,Thevenet2015} showing that the number of ejected electrons reaches a maximum for a density gradient on the order of $\lambda_0/10$ \cite{Thevenet2015}. We also note that at long gradients $L>2$, the ejected charge increases again, see the green circle in fig.~\ref{scan}a). However, we will not focus on the case of longer gradients as we find the ejection mechanism to be quite different from the short gradient case.

In the next section, we will explain the physical mechanism for electron ejection in the regime where $L \ll 1$. In particular, we will show that the plasma fields due to charge separation at the target surface are key to accelerating and ejecting electrons. 

\section{Insights into the mechanism of electron ejection}\label{secscenario}

\begin{figure}[h]
\includegraphics[width=\columnwidth]{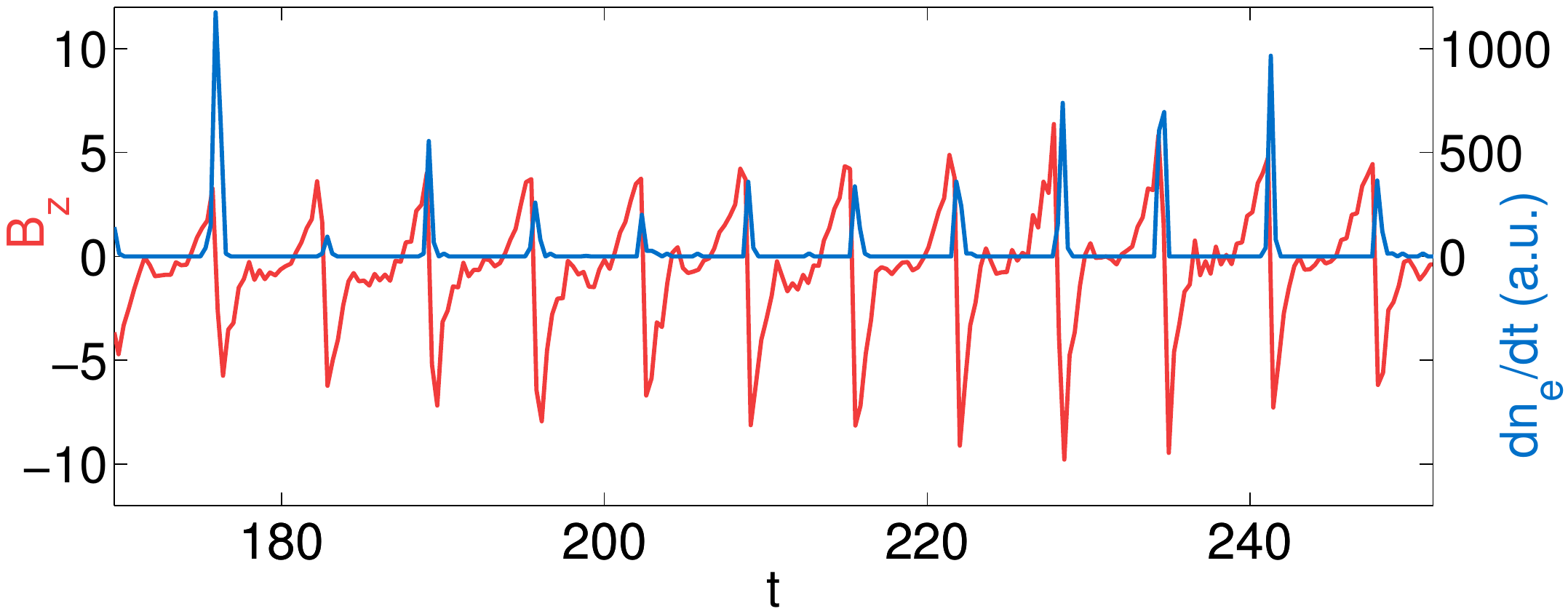}
\caption{1D PIC simulation result for $a_0=5$ and $L=0.55$. Red line: reflected magnetic field at the plasma edge ($x_e=-0.2$) versus time (a Fourier filter was applied to remove the incident field). Blue line: density of ejected electrons versus time at the same position.}
\label{bunches}
\end{figure}

Let us now focus on the details of this ejection mechanism. Figure~\ref{bunches} shows the reflected field and the density of ejected electrons at the plasma edge ($x_e=-0.2$) versus time for an optimal case $a_0=5$ and $L=0.55$. Electrons are ejected out of the plasma in the form of a train of attosecond bunches, which are then injected inside the reflected laser pulse at a precise phase, as observed in ref~\cite{Naumova2004a,Tian2012}. One can clearly see that they leave the plasma at a phase corresponding to a zero of the electric field. Note also that the reflected electric field is extremely distorted and the sharp peaks indicate a rich harmonic content, as expected from high harmonic generation in the ROM mechanism~\cite{Thaury2010}. 

To simplify the problem further, we now focus on the generation of a single attosecond bunch of electrons during a single optical period. We performed a simulation with a square temporal profile for the incident pulse. In this case, the incident fields are simply written as
\[
\begin{array}{ccc}
\bm{E_L}=a_0\sin(t-x)\bm{e_y} & & \bm{B_L}=a_0\sin(t-x)\bm{e_z}
\end{array}
\]
and the initial electron/ion speed is $\bm{v}=-\sin\theta\bm{e_y}$. For $E_L>0$, the electric force accelerates electrons in the $-y$ direction, allowing them to reach speed $v_y \gtrsim -1$ as soon as $a_0>1$~\cite{Gonoskov2011}. The $\bm{v\times B_L}$ force is then directed towards the $+x$ direction, so that it pushes electrons which concentrate and form a sharp density peak into the plasma. 

\begin{figure}[t]
\includegraphics[width=\columnwidth]{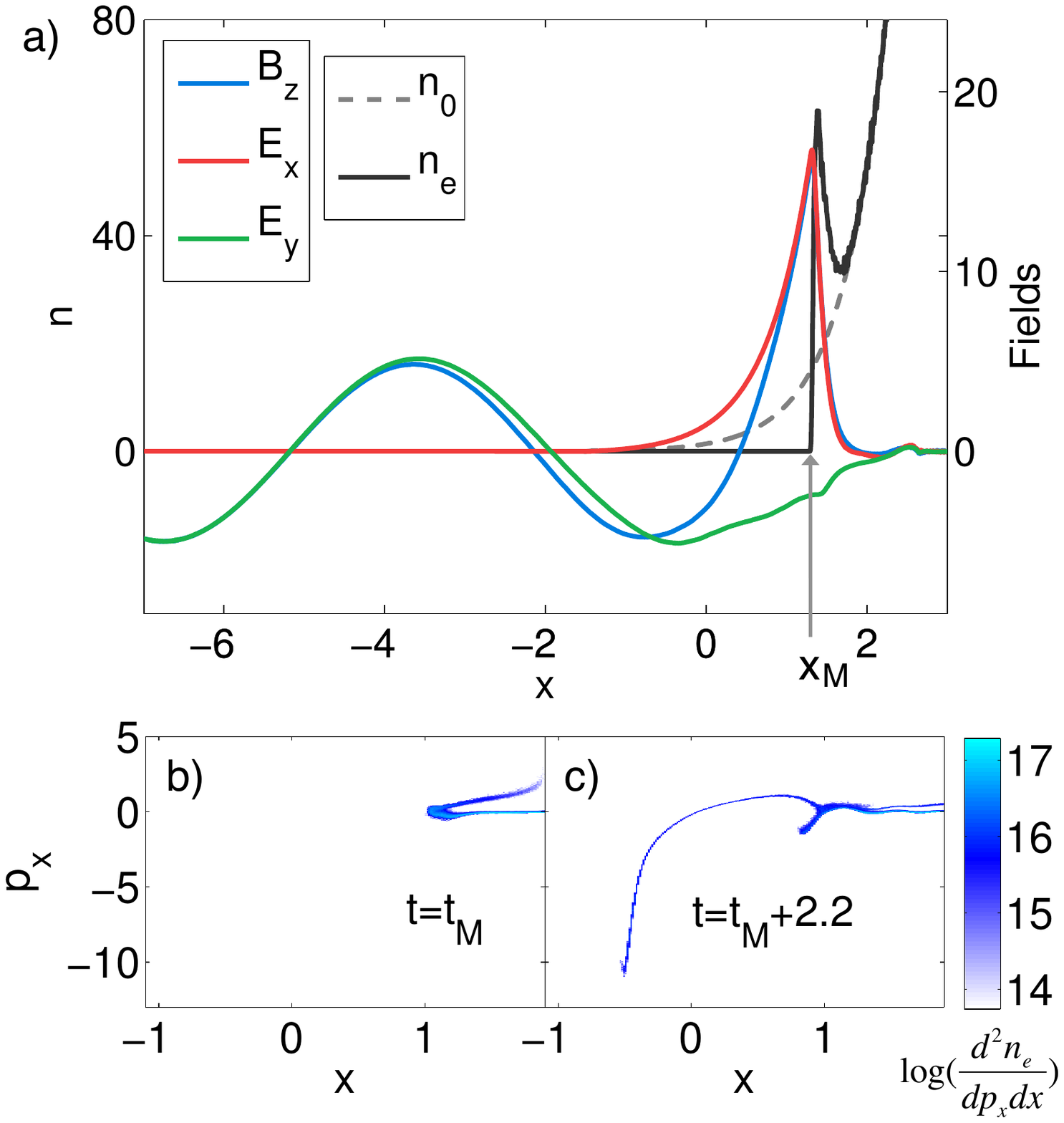}
\caption{a) Density and fields at maximal depth ($t=t_M=14.7$). The black and dashed grey line stand for the electron and ion densities respectively. Color lines are magnetic field $B_z$ (blue), electrostatic field $E_x$ (red) and electric field $E_y$ (green) -- b) and c) Phase space ($x$,$p_x$) at $t=t_M$ and $t=t_M+2.2$ respectively.}
\label{phase}
\end{figure}

This is shown in fig.~\ref{phase}a), where a snapshot of the electron density and the fields is plotted as a function of position $x$. The sharp density peak reaches a maximum depth $x_M$ at $t_M$, after which it travels back towards $-x$. Fig.~\ref{phase}a) shows the density at $t=t_M$, for which all the electrons initially at $x<x_M$ are gathered in the peak. The motion of the density peak during one laser period can be described as a push-pull mechanism.

\emph{(i) Push phase:} During the first half-period, the incident laser field pushes electrons inside the plasma while the ions are not displaced, which builds up an electrostatic plasma field $E_p$ along $x$. Because the plasma is drifting, this charge separation also induces plasma currents which are the source of magnetostatic  plasma field $B_p$. 

At $t=t_M$, the peak reaches an extremum position with $p=0$. At this point, the forces due to the incident wave, the reflected wave and the plasma fields cancel out along the $x$ direction. The maximum peak position can be derived from this balance, as presented in~\cite{Gonoskov2011} for a step-like density profile. It is hereafter extended to the case of a density gradient. 

We first assume that all electrons with initial position $\chi<x$ concentrate in the density peak at $x$. Therefore, the plasma electrostatic field can be obtained from Maxwell-Gauss equation:
\begin{equation}
\frac{\partial E_p}{\partial x} = \frac{1}{\cos^2\theta}(n_i-n_e) \\
\end{equation}
by integrating between $x=-\infty$ and $x$, we obtain 
\begin{equation}\label{eqEp}
\bm{E_{p}}=\frac{1}{\cos^3\theta}Le^{x/L}\bm{e_x}
\end{equation}
 Similarly, integrating Maxwell-Ampere's equation gives $\bm{B_{p}}=\frac{\sin\theta}{\cos^3\theta}Le^{x/L}\bm{e_z}$. At the extremum position $x_M$, the total force along $x$ cancels: $v_y(B_L+B_p)+E_p=0$. In addition, at $x_M$, the laser reflects off the surface and we assume that the boundary conditions of a perfect conductor can be applied for the incident and reflected laser fields, so that at $E_L(x_M,t_M)=0$ and $B_L(x_M,t_M)=2a_0$. We further assume that the electron speed at $t=t_M$ is $\bm{v}=(0,-1)$. The balance equation along $x$ reduces to
\begin{equation}
2 a_0 + \frac{\sin\theta }{\cos^3\theta}Le^{x_M/L}-\frac{1}{\cos^3\theta}Le^{x_M/L} =0
\end{equation}
Solving for $x_M$, we obtain
\begin{equation}
x_M = L\log \bigg[ \frac{2a_0\cos^3\theta}{L(1-\sin\theta)} \bigg].
\end{equation}
This equation gives an estimation of the surface position at maximum depth which fits within less than $20\%$ error when compared to the PIC simulations. At $t_M$, the target surface is similar to a plasma capacitor in which electrons are able to gain energy. The maximum fields of this plasma capacitor can be expressed as 
\begin{eqnarray}\label{eqfields}
E_p(t_M,x_M)=\frac{2a_0}{(1-\sin\theta)}\\
B_p(t_M,x_M)=\frac{2a_0\sin\theta}{(1-\sin\theta)}
\end{eqnarray}
Clearly, the plasma capacitor has larger fields for large $a_0$ and incidence angle approaching grazing incidence. Also note that for large angles $\theta$, the plasma fields are larger than the laser fields. From the electric field, one can estimate the maximum energy gain in this plasma capacitor: $\Delta\gamma\simeq E_p(x_M)L\propto a_0L$. 

\emph{(ii) Pull phase:} During the following half-period, the laser field changes sign so that the $v_yB_L$ force now pulls electrons towards vacuum and breaks the balance of the force along $x$. The electron peak is then accelerated towards vacuum ($x<0$), and radiates an attosecond electromagnetic bunch~\cite{Dromey2006}. A small fraction ($<1\%$) of electrons in the density peak escapes the plasma and travels along the reflected pulse. This is shown in fig.~\ref{phase}b) and c), where the phase space is represented at $t=t_M$ and $t>t_M$ respectively. For $t>t_M$, one can see a jet of electrons traveling towards vacuum, $x<0$ and $p_x<0$. 

\begin{figure}
\includegraphics[width=\columnwidth]{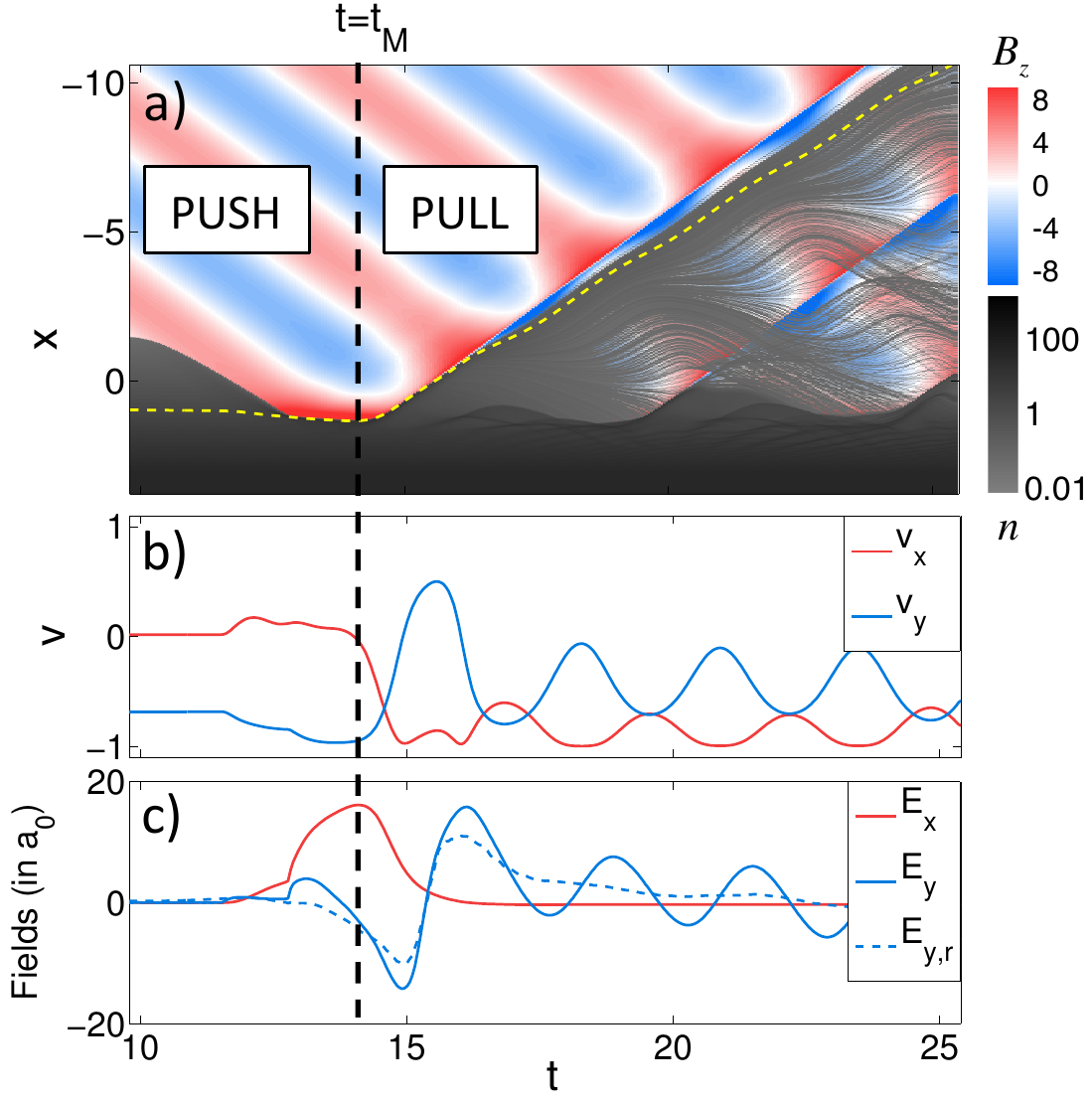}
\caption{Dynamics of an ejected electron for parameters $a_0=5$ $L=0.55$ a) Greyscale: electron density $n_e(x,t)$. Color scale: magnetic field $B_z(x,t)$. The yellow dashed line represents the trajectory of an ejected electron. b) Velocity components of the ejected electron. c) Electric fields seen by the ejected electron along its trajectory.}
\label{mechanism}
\end{figure}

The dynamics of such electrons is detailed in fig.~\ref{mechanism}. Panel a) shows the magnetic field $B_z$ (color scale) and the electron density $n_e$ (grey scale) versus time and space.  The electron trajectory is plotted as a dashed yellow line in fig.~\ref{mechanism}a). It originates from deep inside the plasma (around $x=x_M$) and is released in the plasma capacitor at $t=t_M$.

The electron velocity components $v_x$ and $v_y$ are plotted in panel b). The electron drifts with initial speed $v_y=-\sin\theta$ and at $t_M$, as the laser field changes sign, it is strongly accelerated towards vacuum in a fraction of an optical cycle (typically in $T_0/10$). It then propagates with $v_x \gtrsim -1$ and oscillates in the interference pattern created by the incident and reflected fields, while closely remaining in phase with the reflected field. Panel c) shows the electric and magnetic fields along the electron trajectory. The initial energy gain is given by the electrostatic field $E_x$ at $t \geq t_M$ until it leaves the plasma. The electric field $E_y$ seen by the electron is the superposition of fast oscillations in the incident field and a slow phase shift in the reflected field. The reflected field along the electron trajectory is represented by the dashed line. 

Figure~\ref{mechanism} gives insights on the ejection mechanism: (i) electrons are accelerated toward vacuum by the electrostatic field $E_p$ of the plasma capacitor, (ii) electrons are then able to stay in phase in the reflected field and gain additional energy from the transverse laser field $E_y$ over a long distance via vacuum laser acceleration~\cite{Thevenet2015}. After $t_M$, fig.~\ref{mechanism}b-c) show that the incident field has relatively little effect other than provoking an oscillation of the transverse velocity.

To confirm that this scenario is valid for all ejected electrons, we compute the work of the electric fields along the trajectories of ejected electrons:
\begin{equation}
 \Gamma_{x}=-\int_0^tE_{p}v_{x}dt, \;\;\;\;  \Gamma_{y}=-\int_0^tE_{L}v_{y}dt
 \end{equation}
where $\Gamma_x$ and $\Gamma_y$ represents the energy gain in the plasma field and laser field respectively.

\begin{figure}
\includegraphics[width=\columnwidth]{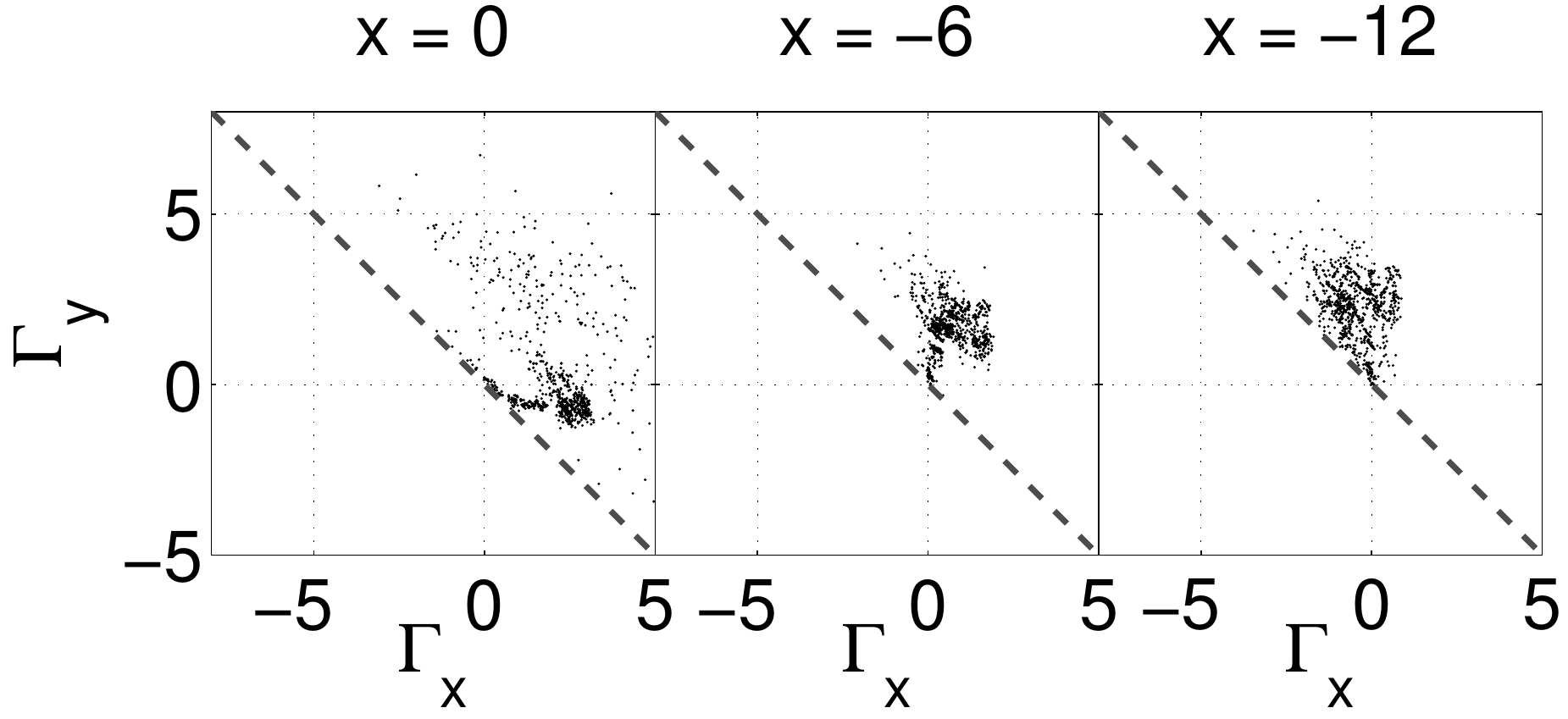}
\caption{Work of the electric fields calculated when electrons cross different planes $x=0$, $x=-6$ and $x=-12$ (simulation parameters are $a_0=10$ $L=0.55$). The $\Gamma_x$ represents the energy gain in the plasma field while the $\Gamma_y$ represents the energy gain in the transverse laser field.}
\label{work}
\end{figure}

The result for a large number of electrons is shown in fig.~\ref{work} for the case with $a_0=10$ and the optimal gradient $L=0.55$. For $x=0$, \textit{i.e.} around the plasma edge, the energy gain is dominated by $\Gamma_x$, which means that it is due to the plasma field $E_p$. On the other hand, as the electrons move away from the plasma surface ($x=-6$), the energy gain is dominated by $\Gamma_y$, i.e. there is additional energy gain from the laser field, via vacuum laser acceleration. Finally, at $x=-12$, the average $\Gamma_x$ is negative, meaning that the recall force from the non-neutral plasma surface is decelerating electrons, an effect which is exaggerated in 1D as discussed previously.

\begin{figure}
\includegraphics[width=\columnwidth]{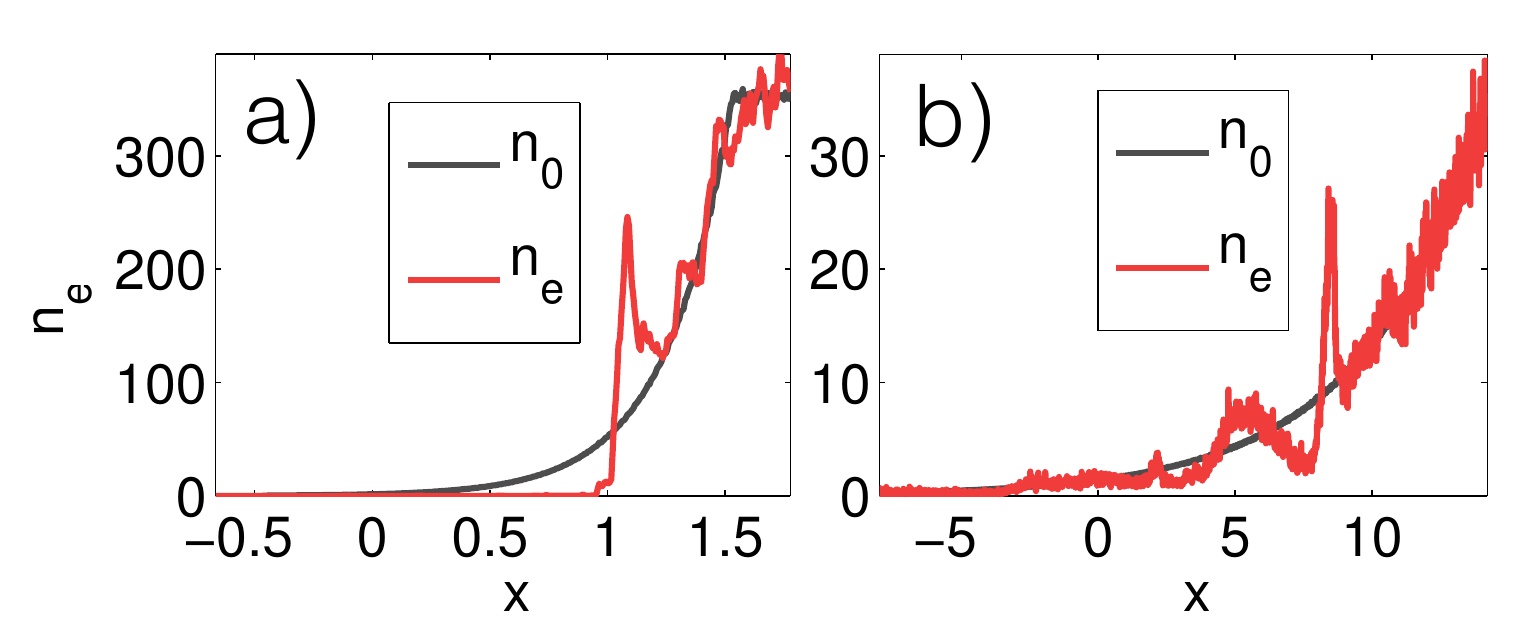}
\caption{a) Electron density at $t=0$ (black) and $t=t_M$ for $a_0=8$ and $L=.55$. b) Same as a) for $L=4.4$. }
\label{potential}
\end{figure}

Finally, when the hypothesis $L < 1$ is broken, in particular when $L \gg 1$, this acceleration and ejection scenario is not valid anymore. Indeed, the formation of a large density peak does not occur for larger gradients,  as shown in fig.~\ref{potential}a) and b), where the electron density profile is plotted at $t=t_M$ for a short and a long gradient respectively. It is readily seen that for the long gradient case, the density in the peak is ten times lower and the width of the peak is also much larger. Consequently, the plasma capacitor does not form and electrons cannot be accelerated efficiently by the plasma field. 

\section{Simplified 1D model of electron ejection}\label{secmodel}

In the previous discussion, we found that the energy gain in the plasma capacitor scales as $\propto a_0L$, explaining qualitatively why the ejected charge increases with $L$ and $a_0$, in the limit of small gradients. However, a more quantitative model would be useful for estimating the ejected charge and typical ejected energy. Several models have been developed for high-harmonics generation~\cite{Lichters1996,Baeva2006,Vincenti2014a}, surface motion~\cite{Sanz2012,Debayle2015} or electron jets inside the plasma~\cite{Ruhl1996}. Yet, none of them is suitable for describing backward electron acceleration under oblique incidence with a density gradient. 
In particular, the model developed by Gonoskov in Ref.~\cite{Gonoskov2011} describes very well the peak dynamics at very large intensity ($a_0 \gg 1$) under oblique incidence and for a step-like density profile. However, the density peak is assumed to travel at the velocity of light $v=1$, \textit{i.e.} infinite energy, and the model cannot be used to solve the equations of motion.

We now propose a simple numerical model to illustrate the ejection process during one optical cycle. The incident laser wave is approximated by a monochromatic plane wave with $a_0>1$. Ions are immobile. We assume $L \ll 1$ so that electrons are gathered in a density peak of width $d \ll L$. In the boosted frame, the ion density profile reads $n_i(x)=n_0(x) = e^{x/L}/\cos\theta$. We consider an electron in the density peak, and describe its motion starting from $t=t_M$. First, the equations of motion for an electron in the density peak are derived. Second, we find the appropriate initial conditions. Third, this set of equations is solved numerically and compared with PIC simulations results.

Electrons are driven by (i) electromagnetic fields and (ii) plasma fields. We describe the motion of electrons in the density peak during the pull phase, during which the reflected field is generated. The incident laser electric field is written $\bm{E_{L,i}}=-a_0\sin(t-x+\phi_{M,i})\bm{e_y}$, where the phase $\phi_{M,i}=x_M-t_M$ is chosen so that the laser field changes sign at $(t_M,x_M)$. For the reflected field, we neglect the harmonic content of the field and simply write $\bm{E_{L,r}}=a_0\sin(t+x+\phi_{M,r})\bm{e_y}$ with  $\phi_{M,r}=-x_M-t_M$.

\begin{figure}
\includegraphics[width=\columnwidth]{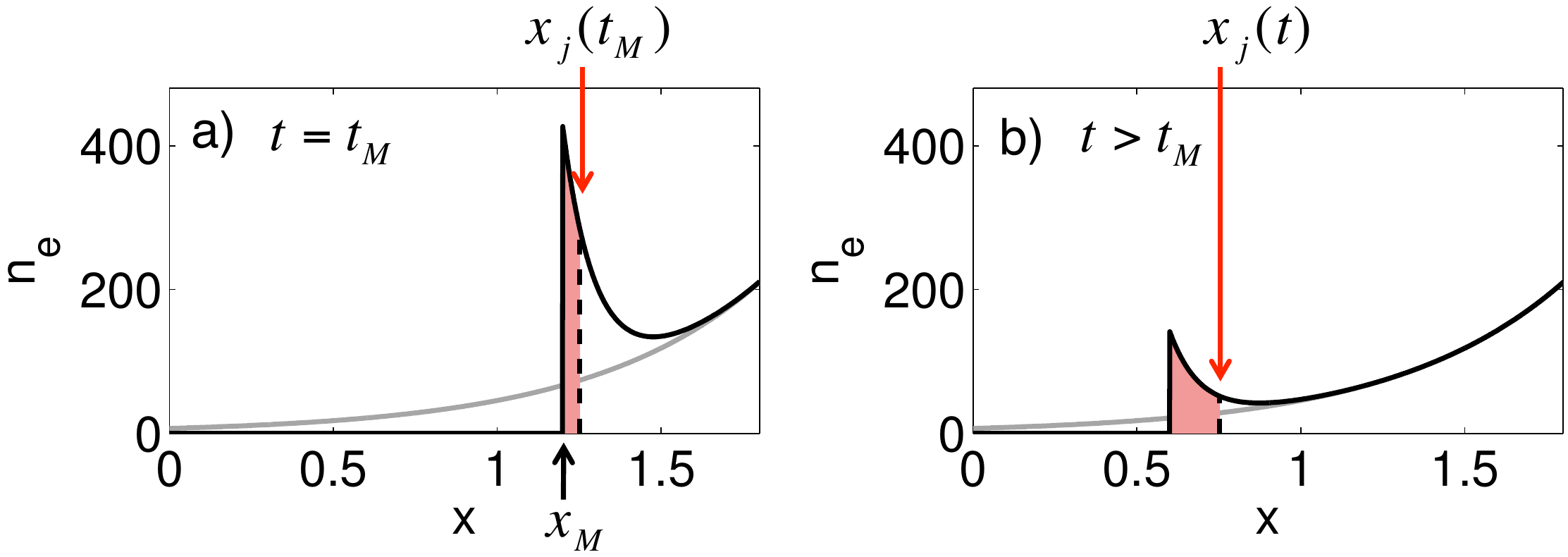}
\caption{a) Sketch of the electron density at $t=t_M$. The black dotted line stands for $x_{j0}$, the position of electron $j$ at $t=t_M$, the initial time for our model. The red area shows the initial charge on the left of electron $j$. b) Same as a) for $t>t_M$. In agreement with the hypothesis of no trajectory crossing, the charge on the left of electron $j$ is conserved along time: the surface of the red area is the same for all $t\geq t_M$.}
\label{peak}
\end{figure}

Concerning the plasma field, it is crucial to include electron screening in the density peak in order to be able to model the ejected charge. Indeed, when an electron $j$ is located on the front edge of the peak ($x_j=x_M$), it experiences the full plasma field $E_p$ (see fig.~\ref{peak}). On the contrary, an electron located at $x_j>x_M$ experiences a screened plasma field $E_p-E_s$, where $E_s$ stands for the electronic screening field, and is less likely to escape the plasma. As seen before, the plasma fields at the position of electron $j$, $x_j(t)$ can be obtained by integrating Maxwell-Gauss's equation, giving
\begin{equation*}
E_p-E_s=\frac{1}{\cos^2\theta}\left(\int_{-\infty}^{x_{j}(t)}n_i(x)dx-\int_{-\infty}^{x_{j}(t)}n_e(x)dx\right)
\end{equation*}
The first term $E_p$ is evidently the unscreened plasma electric field from eq.~\ref{eqEp}, while the second term is the screening electric field coming from electrons in the density peak. 


Since the shape of the density peak cannot be calculated analytically, the second integral cannot be evaluated easily. Therefore, the screening field $E_s$ is derived by assuming that there is no trajectory crossing \cite{Brunel1987}: if electrons $j$ and $k$ in the density peak verify $x_j(t_M)<x_k(t_M)$, then $x_j(t)<x_k(t)$ at any time $t \geq t_M$. With this assumption, the number of electrons on the left of electron $j$, \textit{i.e.} $x<x_j(t)$, is constant along time, see figure~\ref{peak}. 

Hence, the integral of the electron contribution in the Maxwell-Gauss equation is conserved,

\begin{eqnarray}
E_{s} & = & \int_{-\infty}^{x_j(t)}\frac{n_e(x,t)}{\cos^2\theta} dx \nonumber \\
        & = & \int_{-\infty}^{x_j(t_M)}\frac{n_e(x,t_M)}{\cos^2\theta}dx \nonumber \\
        & = & \sigma_j
\end{eqnarray}

where $\sigma_j$ is the initial surface charge on the left of electron $j$. Therefore, the screening field $E_s$ is simply determined by the surface charge $\sigma_j$, and there is no need to know the details of the shape of the density peak. This electronic surface charge screens the plasma field $E_p$ and reduces the acceleration of electrons. Note that this screening field is constant in time and therefore it has a considerable effect on the electron trajectories. As the electron peak moves along $x$ and $y$, it radiates a magnetic field $B_s$ through Maxwell-Ampere equation which is reponsible for the reflected field. Neglecting the high harmonics, this radiation comes down to a monochromatic plane wave that we include in the equation of motion. Finally, taking these effects into account, the equation of motion for electron $j$ reads

\begin{eqnarray}\label{eqmotion}
\frac{d\bm{p_j} }{dt} 
& = & +a_0\sin(t-x_j(t)+\phi_{M,i}) \Big[ \bm{e_y} + \bm{v_j(t)}\times\bm{e_z} \Big] \nonumber \\
&  & -a_0\sin(t+x_j(t)+\phi_{M,r}) \Big[ \bm{e_y} - \bm{v_j(t)}\times\bm{e_z} \Big] \nonumber \\
& & - \frac{L}{\cos^3\theta}e^{x_j(t)/L} \Big[ \bm{e_x} + \sin\theta \bm{v_j(t)}\times\bm{e_z} \Big] \nonumber \\
& & + \sigma_j\bm{e_x}
\end{eqnarray}
The first and second lines are for the incident and reflected waves, the third line is for the plasma capacitor fields and the last line is the screening field. The initial conditions are taken at $t_0=t_M$, when electrons in the density peak are located at $x_M<x<x_M+d$. Since $d<<L$, we assume that all electrons start at $x=x_M$ and use the expression of $x_M$ given above.

\begin{figure}
\includegraphics[width=\columnwidth]{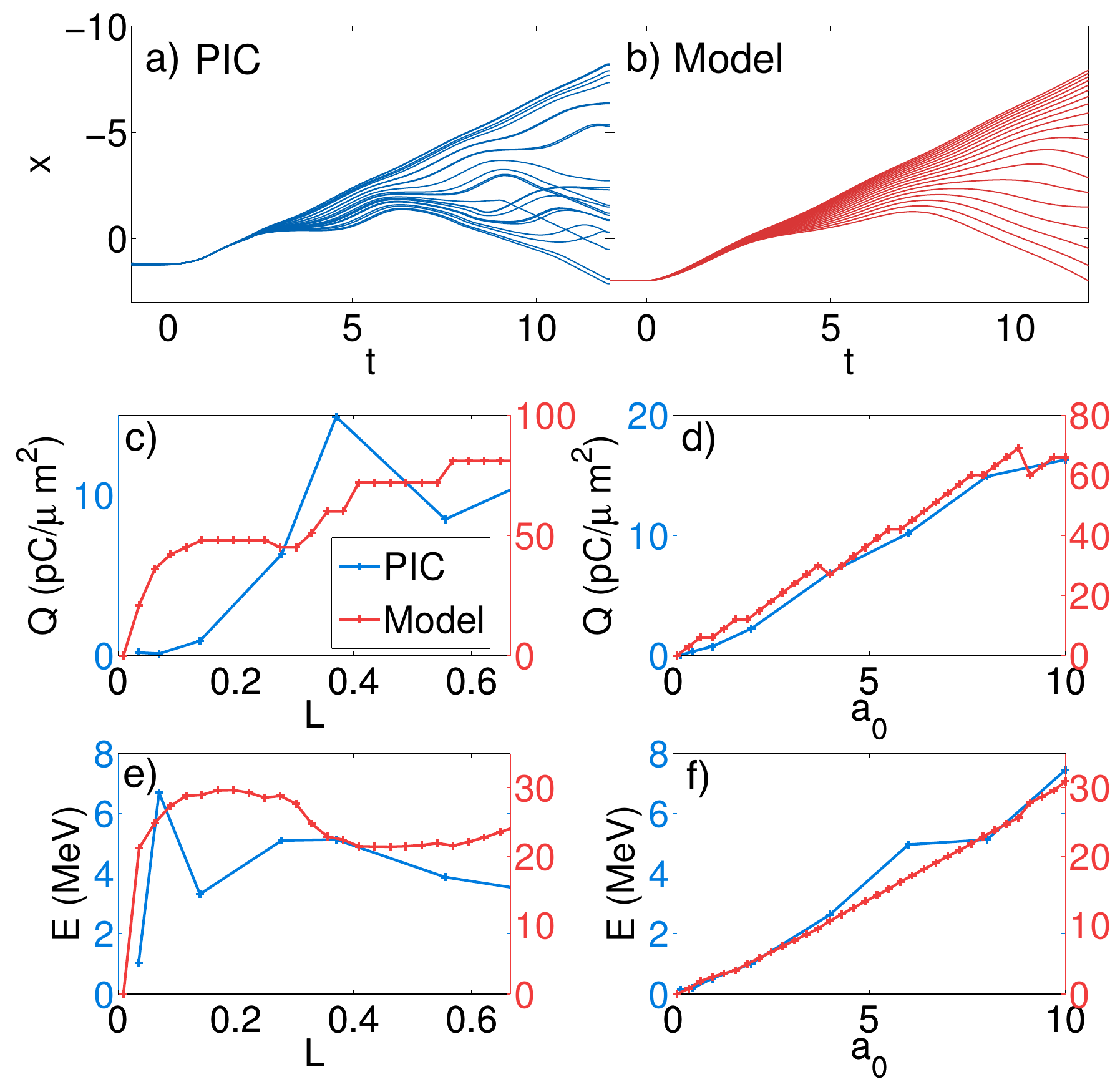}
\caption{a) Electron trajectories from PIC simulation $a_0=8$ $L=.55$. b) Electron trajectories from the model, with the same parameters. In both cases, $t=0$ stands for $t_M$. c) Ejected charge versus $L$ for $a_0=8$ from PIC simulations and our model. d) Ejected charge versus $a_0$ for $L=0.36$ from PIC simulations and our model. e) Same comparisons between the PIC simulations and the model but for the average energy of the ejected electrons.}
\label{model}
\end{figure}

The initial momentum of electrons is: $p_{x0}=0$ because at $t_M$ the peak position is extremal. The transverse momentum $p_y$ is derived from the conservation of the canonical momentum $P_y=p_y-a_y=P_{y0}$. The peak reaches its maximum depth when the incident field changes sign, \textit{i.e.} $a(t_M,x_M)=a_0$. The initial conditions are the same for all electrons and read
\[
\left \{
\begin{array}{rcl}
x_{0} & = & x_M \\
p_{x0} & = & 0 \\
p_{y0} & = & -\tan\theta-a_0
\end{array}
\right.
\]
Finally, the only difference between electrons $j$ and $k$ is the initial charge on the left side of the electron, \textit{i.e.} the term $\sigma_j$ in equation ~\ref{eqmotion}.
These equations are solved numerically for different values of $\sigma_j$. The ejected charge can be determined by increasing $\sigma_j$ until a threshold value $\sigma_{max}$ above which the electron is not ejected; the ejected charge is then simply $\sigma_{max}$. An example is given on fig.~\ref{model}, where electron trajectories are plotted for $a_0=8$ and $L=0.55$, from a PIC simulation (a) and using the model (b). There is no trajectory crossing in the PIC simulation before $t=10$, which validates our hypothesis. The global dynamics is very well reproduced.  

The following ejection criterion was adopted in the model: an electron is considered to be ejected if $p_x$ is negative during $3$ periods. This criterion is different from the one we adopted for PIC simulations because we assumed no trajectory crossing, which is not valid for a large time range.

Fig.~\ref{model}c) and d) show the ejected charge plotted versus $L$ and $a_0$ respectively. The model reproduces the global trends: the charge increases with $a_0$ and $L$. It overestimates the ejected charge because the ejection criterion is much more stringent for the PIC simulation than for the model.The linear scaling of ejected charge with $a_0$ is well reproduced.  The scaling with the density gradient does not fit as well which can be explained by the fact that as $L\rightarrow1$ , the plasma capacitor model collapses. Remarkably, our simple model also reproduces the trends and the order of magnitude for the ejected electron energy. This is shown in fig. ~\ref{model}e) and f).

To illustrate the role of plasma effects, the model was run with no ion plasma fields (removing the third line in equation \ref{eqmotion}), in the same conditions as fig.~\ref{model}c). The ejected charge never exceeded $1\;\mathrm{pC/\mu m^2}$, which clearly validates the plasma capacitor model for electron ejection. 
When we run the model without the reflected field, we find that the ejected charge increases linearly with the gradient length $L$ instead of saturating at longer gradients. This shows that the reflected field also play a role in the details of the ejection.

\section{2D PIC simulations}\label{sec2D}

We now show the results of 2D PIC simulations in order to confirm the validity of the 1D study. We also use this more realistic 2D geometry to study the correlation between electron ejection and the emission of high harmonics. Indeed, the ejected electrons are part of the density peak that oscillates in the plasma. When this density peak is accelerated toward vacuum, it radiates the reflected field and its high harmonics via the ROM mechanism. Thus, the ejected electrons and the ones that radiate the high harmonics should be closely related and we expect high harmonic emission to be correlated to the ejected charge. 

The 2D simulations are run in the laboratory frame and the normalization is done as follows
\[
\begin{array}{rclcrcl}
t & \equiv & \omega_0 t & & x & \equiv & k_0x \\
v & \equiv & v/c & & p & \equiv & p/m_ec
\end{array}
\]
with $\omega_0$ and $k_0$ the laser angular frequency and wave vector in the lab frame respectively.

Modeling simultaneously electron ejection and high-harmonic generation in a 2D PIC simulation is challenging. First for HHG, for one should use a spatial mesh of the order of the Debye length $\Delta x \simeq \lambda_{De}=0.036 \sqrt{\frac{T_{e[\kev]}}{n_{e,max}}}$, where $T_e$ is the plasma electron temperature in $\kev$ and $n_{e,max}$ is the maximum electron density. In our case, this condition reads $\Delta x = 10^{-3}$ for an electron temperature $T_e=0.1\kev$ and bulk density $n_{0}=250$.

On the other hand, ejected electrons should be detected far from the plasma surface, at least at a distance larger than the spot size, and ideally larger than the Rayleigh length, which requires a large 2D simulation box size $D=600$. Satisfying both condition leads to a simulation box with $10^6\times10^6$ cells with $10^{11}$ particles, which is far too large for this study. Hence, we released these conditions and ran the simulations with $n_{0}=100$ and $T_e=0.1\kev$. The simulation results converge when using a spatial mesh as small as $\Delta x=0.021$. The simulation box size was $D_x\times D_y=530\times200$.


A $800\nm$, laser pulse impinges the solid-density plasma with a $\theta=45^{\circ}$ incidence angle. The pulse duration is $=25\fs$, its spot size is $3.4\mic$ FWHM and its amplitude is $a_0=3$. We performed simulations for the following gradient scale lengths: $L=0.2, 0.8, 1.6, 3.1, 6.3$. Ejected electrons are detected with two electron probes. The first one is parallel to the plasma surface and located at a distance of $160$ away from the plasma surface to record electrons emitted around the specular direction. The second one is perpendicular to the plasma surface and $160$ from the reflection point, to record electrons ejected along the plasma surface.

The 2D simulations reproduce the main phenomena depicted in the 1D PIC simulations: at each optical cycle, electrons are pushed and form a sharp density peak. This gives rise to a plasma capacitor in which electrons gain energy and are ejected. Jets of electrons are ejected at precise phases of the reflected laser field (at zeros of the electric field) and further propagate first in the interference pattern and then in the reflected field.

\begin{figure}
\includegraphics[width=\columnwidth]{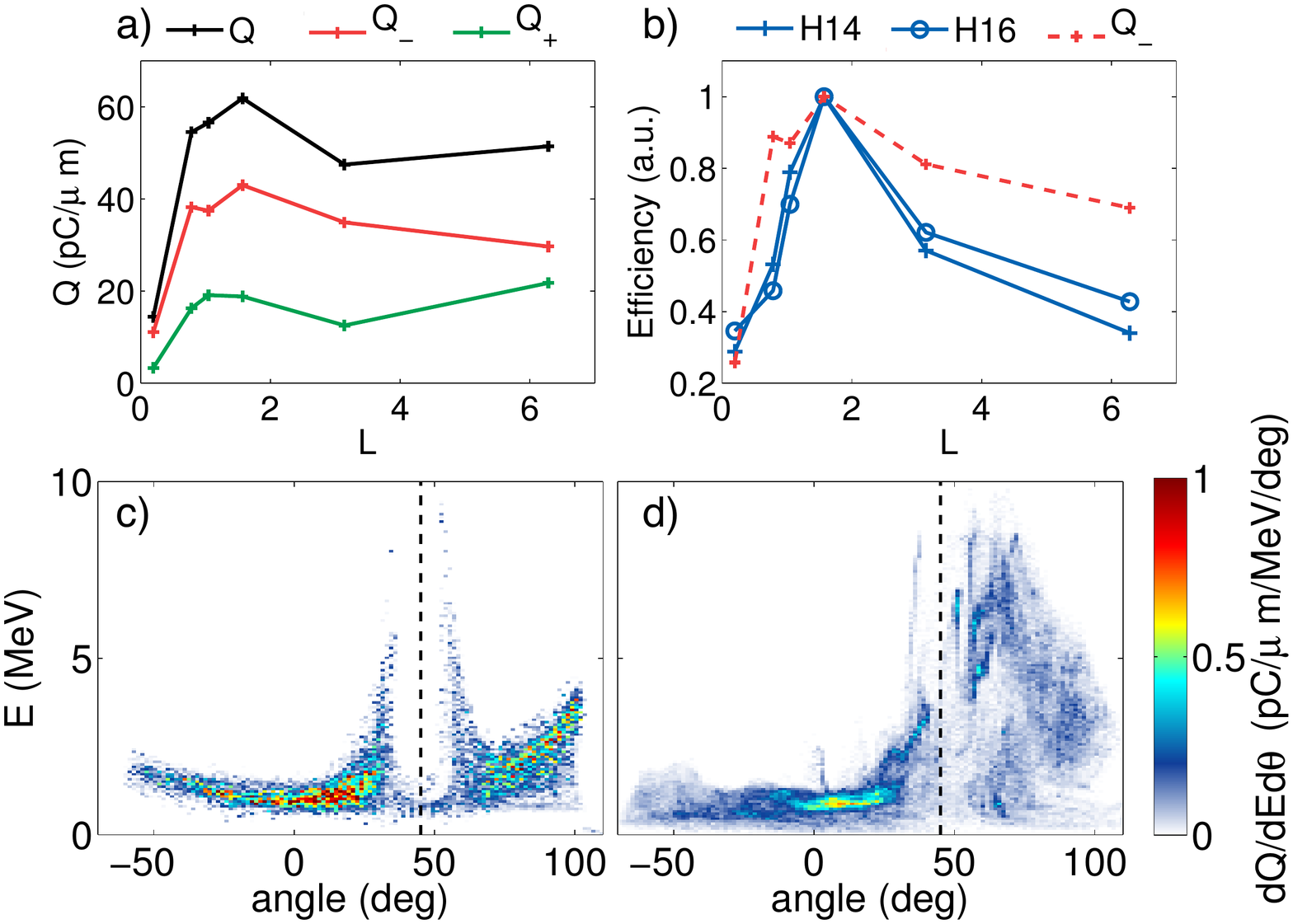}
\caption{Results of 2D PIC simulations for $a_0=3$ and various gradient lengths. a) Ejected charge as a function of the gradient length. Electrons are sorted depending on their final emission angle: $Q_+$ is the charge for electrons with $\theta>45^\circ$, and $Q_-$ is the charge for electrons with $\theta<45^\circ$, so that $Q=Q_-+Q_+$. b) Correlated emission of harmonics and charge $Q_-$. c) and d) Final angle-energy distribution for $L=0.8$ and $L=6.3$ respectively. The vertical dashed black line shows the specular direction.}
\label{2dscan}
\end{figure}

More qualitative results are shown in fig.~\ref{2dscan}. The black line on panel a) shows the total ejected charge (\textit{i.e.} on both electron probes) as a function of the gradient length. As previously, the ejected charge increases with the gradient scale length until it reaches a maximum for $L\simeq 1.7$. This qualitatively confirms the observations from the 1D PIC simulations, although the optimal gradient length $L_{max}$ is longer: $\simeq 1.6$ instead of $\simeq 0.5$ in 1D.

Previous experiments showed that for short gradients, electrons are emitted between the normal and specular direction, so that the electron beam is not symmetric around the specular direction. This asymmetry can be explained by the dynamics of electrons in the reflected field while they undergo vacuum laser acceleration~\cite{Thevenet2015}.


To take into account this asymmetric emission, the ejected electrons are sorted as a function of their final emission angle. We define $Q_-$, the ejected charge of electrons with angles $<45^\circ$ (\textit{i.e.} between the normal and specular direction) and $Q_+$ as the charge for electrons emitted with angles $>45^\circ$ (\textit{i.e.} between the specular and grazing directions). Fig.~\ref{2dscan}a) shows that $Q_-$ decreases for gradients above $L\simeq 1.7$, while on the contrary $Q_+$ increases for long gradients. The behavior of $Q_-$ is consistent with the plasma capacitor scenario while the opposite behavior of $Q_+$ indicates a different ejection mechanism which dominates for longer gradients. These two regimes give rise to significantly different angular distributions, as seen in fig.~\ref{2dscan}c) and d). Panel c) shows the angle-energy distribution in the case of a short gradient ($L=0.8$) for which the majority of electrons are ejected with angle $\theta<45^\circ$. In this case, the ponderomotive force digs a hole close to the specular direction as it pushes electrons away from the laser pulse~\cite{Thevenet2015}. Panel d) shows the case of a longer gradient ($L=6.3$): more electrons are ejected along the target, indicating a different emission process.



Finally, fig.~\ref{2dscan}b) represents the harmonics signal as a function of the gradient scale length. The harmonics efficiency peaks for $L\simeq 1.8$ and decreases for longer gradients, and follows identical variations as the charge $Q_-$. This confirms that electron ejection in the plasma capacitor (represented by $Q_-$) and the high-harmonics generation via the ROM mechanism are correlated. 


\section{Conclusion}

We have studied the process of electron ejection, or backward emission, when a solid target with a sharp density gradient is irradiated by an ultra-intense laser at oblique incidence. An extensive numerical study provides insights into the ejection mechanism in the case of short gradients $L<\lambda_0$. In this case, the laser pushes electrons inside the density gradient and sets up a high density peak of electrons associated to a large charge-separation field, akin to a plasma capacitor. Electrons can gain enough energy in this plasma field to be expelled from the plasma surface into the vacuum and injected into the reflected field in which they eventually gain additional energy. A 1D model of such electron ejection was developed and found to reproduce all the trends of the PIC simulations. Finally, our 1D study is completed by 2D PIC simulations validating the ejection mechanism and showing that electron ejection is closely correlated to harmonic emission via the relativistic oscillating mirror mechanism. 

Electrons accelerated at these plasma surfaces have very specific characteristics: they are emitted with MeV energy, at a given phase of the field and with an attosecond bunch duration. Therefore, these plasma mirrors are ideal for injecting electrons into a laser field and accelerating them via vacuum laser acceleration, as demonstrated in~\cite{Thevenet2015}. All these processes result in the generation of a source of MeV electrons with high charge, large divergence and femtosecond duration. Using few cycle laser pulses could permit the generation of a single sub-femtosecond electron bunches, which in conjunction with the attosecond harmonics pulse, could be used to probe matter in extreme conditions. 



This work was funded by the Agence Nationale pour la Recherche (under contract  ANR-14-CE32-0011-03 APERO) and the European Research Council (under contract No. 306708, ERC Starting Grant FEMTOELEC). Access was granted to the HPC resources of CINES and CCRT under allocation 2015-056057 made by GENCI. Simulations were run using EPOCH, which was developed as part of the UK EPSRC funded projects EP/G054940/1.


\end{document}